\documentclass[superscriptaddress,
twocolumn,showpacs,preprintnumbers,amsmath,amssymb]{revtex4}
\usepackage[dvips]{graphicx}
\usepackage{epsf}
\usepackage{epsfig}
\usepackage{latexsym}
\usepackage{amssymb}
\usepackage{amsfonts,amsbsy}
\usepackage{amsmath}
\usepackage{dcolumn}
\usepackage{bm}
\usepackage{pifont}
\begin{document}
\title{Collective effects in intra-cellular molecular motor transport:\\
coordination, cooperation and competetion \footnote{Section \ref{seckif} is 
based on original work \cite{nosc,greulich} carried out in collaboration with K. Nishinari, Y. Okada, A. Schadschneider, P. Greulich and A. Garai. Section \ref{secribo} is based on original work \cite{basuchow} done in collaboration with A. Basu.}}
\author{Debashish Chowdhury}
\affiliation{%
Department of Physics, Indian Institute of Technology,
Kanpur 208016, India.
}%
\date{\today}
\begin{abstract}
Molecular motors do not work in isolation {\it in-vivo}. We highlight 
some of the coordinations, cooperations and competitions that determine 
the collective properties of molecular motors in eukaryotic cells. 
In the context of traffic-like  movement of motors on a track,  
we emphasize the importance of single-motor bio-chemical cycle and 
enzymatic activity on their collective spatio-temporal organisation.
Our modelling strategy is based on a synthesis- the same model describes 
the single-motor mechano-chemistry at sufficiently low densities whereas 
at higher densities it accounts for the collective flow properties 
and the density profiles of the motors. We consider two specific 
examples, namely, traffic of single-headed kinesin motors KIF1A on 
a microtubule track and ribosome traffic on a messenger RNA track.
\end{abstract}
\maketitle
\section{\label{sec1} Introduction}

Molecular motors \cite{schliwa,hacktama} perform crucial functions at 
almost every stage in the life of a cell. The motor-dependent activities 
begin with DNA replication \cite{alberts} which is essential before 
cell division. A motor 
called DNA helicase \cite{levin,lohmann} unzips the double-stranded 
DNA so that another motor, called DNA polymerase \cite{rothwell} can 
synthesize two copies of the DNA using the two single strands as 
templates.  Protein synthesis \cite{alberts}, one of the crucial 
activities within a cell, is accomplished by some other motors at 
different stages of operation, the most notable among these are RNA 
polymerase \cite{gelles} and ribosomes \cite{ramakrishnan}. Normally, 
the proteins are synthesized near the cell center and need to be 
transported to appropriate locations some of which are far from the 
cell center. Besides, some cargoes are also transported from the cell 
periphery to the cell center. 
Most of these transportations of cargoes (often in appropriately 
packaged form) in the cytoplasm are carried out by yet another set 
of molecular motors which are collectively referred to as cytoskeletal 
motors \cite{howard}. 

The molecular motors at the cellular and subcellular levels can 
be grouped together is several different ways. One possible way of 
grouping these is based on the nature of the tracks on which the 
motors move- one group moves on filamentary proteins whereas the 
other moves on nucleic acid strands. Among the cytoskeletal motors, 
which walk on filamentary proteins, kinesin and dynein move on 
microtubules (MT) whereas myosins move on actin filaments. Some 
motors are processive, i.e., capable of long walk without getting 
detached from the track. Kinesins and dyneins are not only 
processive but can carry molecular cargo; therefore, these are 
functionally similar to ``porters''.

Molecular motors do not work in isolation {\it in-vivo}. Wide range 
of biophysical phenomena observed at sub-cellular and cellular 
levels are manifestations of different types of collective processes,  
involving molecular motors, at several different levels of biological 
organisation \cite{vermeulen}. The nature of the collective effect 
depends on the situation; it can be the {\it coordination} of different 
parts of a single motor or, on a larger scale, it can be a 
{\it cooperation} or {\it competition} between two different motors. 
These collective effects can manifest as spontaneous oscillations, 
dynamic instabilities, hysteresis, dynamical phase transitions 
\cite{kruse05,vilfan05} and motor traffic jam \cite{polrev}. 
In this paper, we first give examples of different types of 
coordination, cooperation and competition in molecular motor 
transport. Then, we focus on our works on molecular motor traffic 
on two different types of tracks, namely, on MT and on messenger 
RNA (mRNA).

\section{Coordination through elastic coupling}

In this section we consider coordination of elastically coupled 
motor domains of a given motor as well as that of motors which 
are elastically coupled to a common backbone. The structures and 
mechanisms of single cytoskeletal motors have been discussed in 
this proceedings by Ray \cite{ray}, and by Mallik and Gross 
\cite{mallikgross} while those of helicase motors have been 
reviewed by Tuteja and Tuteja \cite{tutejas}.

\subsection{Coordination of different heads of oligomeric motors}

Most of the known common molecular motor proteins are dimeric or, 
in general, oligomeric \cite{mei05}. It is 
generally believed that coordination of the ATPase activities of 
different motor domains is essential for the processivity of a 
given motor. The mechanism of this coordination has been 
investiagted extensively over the last decade using many different 
techniques. 

It is now quite well established that the conventional double-headed 
kinesins follow a hand-over-hand mechanism (exactly similar to the 
steppings used by humans for walking) 
\cite{vale97,hackney04,yildiz05,cross04,carter05,carter06}. 
Similar hand-over-hand mechanism is also believed to govern the 
movement of myosin-V, an unconventional processive myosin
\cite{rock04,sellers06}. 
However, the corresponding mechanism for dimeric cytoplasmic 
dynein is much more complex because the architecture of its motor 
domain is very different from those of kinesin and myosin-V 
\cite{oiwa,mallik}.

Majority of the molecular motors that move on nucleic acid tracks (DNA 
or RNA) are also oligomeric. For example, most DNA helicases are 
either dimeric or hexameric \cite{levin}. Both ``inchworm'' and 
``hand-over-hand'' (also called ``rolling'') mechanisms for the 
coordination of the motor domains of dimeric DNA helicases have 
been considered \cite{lohmann04}. The corresponding mechanism for 
hexameric DNA helicases remains unclear; at least three different 
plausible scenarios have been suggested. The ATPase activity of 
the motor domains in hexameric DNA helicases can run (i) in parallel, 
or (ii) in ordered sequential, or (iii) in random-sequential manner 
\cite{patel}.

\subsection{Collective dynamics of cytoskeletal motors bound to a common elastic backbone}

\begin{figure}[tb]
\begin{center}
\includegraphics[width=0.4\textwidth]{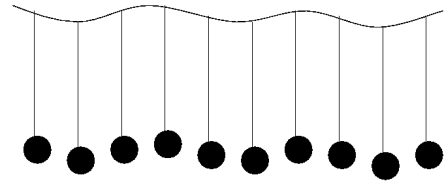}\\
\end{center}
\caption{Schematic description of the model of elastically coupled 
cytoskeletal motors \cite{badoual}; the horizontal curve represents 
the elastic backbone.}
\label{fig-badoual}
\end{figure}

Consider a group of identical motors bound to an elastic backbone as 
shown in fig.\ref{fig-badoual}. It has been shown \cite{badoual} that,  
even if each individual motor is non-processive, such a system of 
elastically coupled motors can move collectively on a filamentary 
track in a processive manner in one direction for a period of time 
and, then, spontaneously reverse its direction of motion.
Such spontaneous oscillations can account for the dynamics of axonemes, 
which are core constituents of eucaryotic cilia, as well as oscillatory 
motions of flight muscles of many insects \cite{kruse05,hilfinger}. 

\section{Two superfamilies of porters: load-sharing versus tug-of-war}

Kinesins and dyneins are both processive (i.e., 
``porters''), but move in opposite directions along MT tracks. Often 
several members of the same superfamily together carry a single cargo. 
On the other hand, if members of both the superfamilies are adsorbed 
simultaneously on the same cargo, they compete against each other 
trying to pull the cargo in opposite directions. Recent progress in 
theoretical modelling of these phenomena have been summarized in this 
proceedings by Lipowsky et al.\cite{lipoetal}

\subsection{Unidirectional transport: load-sharing by members of one superfamily}

Two types of {\it in-vitro} motility assays have been used for studying 
cytoskeletal molecular motors. In particular, in the bead assay the 
filamentary tracks are fixed to a substrate and motors are attached to a 
micron-sized bead (usually made of glass or plastic). The movement of the 
bead in the presence of ATP is monitored using appropriate optical 
micropscopic methods. 
In such situations, each bead is likely to be covered by $N$ motors 
where, in general, $N > 1$. More than one motor is also used for 
transportation of large organelles {\it in-vivo}. This phenomenon of 
unidirectional cooperative cargo transport by more than one motor of 
the same superfamily has been studied theoretically \cite{lipopnas}. 
The average walking 
distance increase with increasing $N$. 
In the face of opposition by external load force, the force is shared 
by the $N$ motors. 

Cooperative pulling of lipid bilayer membranes by kinesin motors can 
generate membrane tubes \cite{roux,koster,leduc}. Such minimal systems 
are adequate to gain insight into the role of motor-mediated interaction 
between the cytoskeleton and organelles in the formation of endoplasmic 
reticulum and Golgi membrane networks.

In case of transport by cytoplasmic dynein motors, if more than one 
motor work together then collectively they can improve the 
performance as porter although individually, while working in isolation, 
they often pause and experience slippage in presence of load force 
\cite{malliketal}. 

\subsection{\label{sec3} Bidirectional transport: tug-of-war between two superfamiles}

In the preceeding section we have mentioned how the a single assembly of elastically 
coupled motors can spontaneously reverse its direction of motion on the 
same track. It is also well known that some motors reverse the direction 
of motion by switching over from one track to another which are oriented 
in anti-parallel fashion. In contrast to these types of reversal of 
direction of motion, we consider in this section those reversals where 
the cargo uses a ``tug-of-war'' between kinesins and dyneins to execute 
bidirectional motion on the same MT track \cite{welte04,gross04}. 
Several possible functional advantages of bidirectional transport have been conjectured \cite{gross04,welte04}. 

Wide varieties of bidirectional cargoes have already been identified so 
far; these include organelles (for example, mitochondria) as well as 
secretaory vesicles and even viruses. If motion in one direction dominates 
overwhelmingly over the other, it becomes extremely difficult to identify 
the movement unambiguously as ``bidirectional'' because of the limitations 
of the spatial and temporal resolutions of the existing techniques of 
imaging.

The main challenge in this context is to understand the mechanisms 
of this bidirectional transport and those which control the duration of 
unidirectional movement in between two successive reversals. This insight 
will also be utilized for therapeutic strategies. For example, the motor 
or the motor-cargo link may be targeted blocking the virus that hijacks 
the motor transport system to travel towards the nucleus. On the other 
hand, a virus executing bidirectional movements can be turned away from 
the outskirts of the nucleus by tilting the balance in favour of the 
kinesins.

\begin{figure}[tb]
\begin{center}
\includegraphics[width=0.4\textwidth]{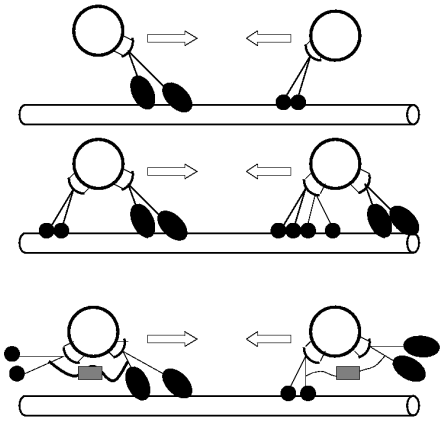}\\
\end{center}
\caption{Schematic description of the three possible mechanisms of 
bidirectional traffic. The three figures from top to bottom 
correspond to the three mechanisms (i), (ii) and (iii) explained 
in the text. The open circle and the rectangular box represent, 
respectively, a micron-size bead and a regulator of bidirectional 
transport.
}
\label{fig-bidir}
\end{figure}

At least three possible mechanisms of bidirectional transport have been 
postulated (see 
fig.\ref{fig-bidir}). (i) One possibility is that either only + end 
directed motors or only - end directed motors are attached to the 
cargo at any given instant of time. Reversal of the direction of movement of the cargo 
is observed when the attached motors are replaced by motors of opposite 
polarity. (ii) The second possible mechanism is the closest to the 
real life ``tug-of-war''; the competion between the motors of opposite 
polarity, which are simultaneously attached to the same cargo and tend 
to walk on the same filament generates a net displacement in a direction 
that is decided by the stronger side. (iii) The third mechanism is 
based on the concept of regulation; although motors of opposite polarity 
are simultaneously attached to the cargo, only one type of motors are 
activated at a time for walking on the track. In this mechanism, the 
reversal of the cargo movement is caused by the regulator when it 
disengages one type of motor and engages motors of the opposite polarity.
For experimentalists, it is a challenge not only to identify the 
regulator, if such a regulator exists, but also to identify the mechanism 
used by the regulator to act as a switch for causing the reversal of 
cargo movement. Dynactin has been identified as a possible candidate 
for the role of such a regulator \cite{gross03,dell03}.

\section{\label{sec4} Transfer from MT networks to actin networks: Park-and-Ride transport system }

The networks of MTs and actin filaments are not disconnected.  
The cytoskeleton is MT-rich near the cell center whereas dense actin filaments
dominate the cytoskeleton near the cell periphery. In recent years 
functional cooperation between the MT-based and actin-based transport 
systems have been discovered \cite{brown,goode,rodionov}.

Several organisms are known to have an intrinsic ability to manipulate 
their skin colour. For example, in fish and frog this change of skin 
colour is caused by the appropriate movement of pigment-containing 
vesicles called melanosome within a special type of cells called 
melanophore. When stimulated, 
melanosomes disperse throughout the cell causing darkening of the 
skin colour. A second stimulus can reverse the process by aggregation 
of the melanosomes near the cell center. During dispersion melanosomes  
are transported, some part of their way towards the cell periphery, by 
kinesin motors. Then, melanosomes switch from the MT-based kinesin 
motors to actin-based myosin motor (unconventional myosin-V which is 
a processive motor) for further distribution around the cell periphery. 
During the reverse process, the melanosomes begin their journey on 
the actin-based transportation network and, somewhere along the way, 
switch over to the MT-based network for continuing their onward 
journey towards the cell center. This switching of the transport 
system is similar to park-and-ride system- 
one drives a car from the suburban areas to the nearest station of the 
urban high-speed mass transit system (railways or metro trains) for 
reaching the central part of a city \cite{maniak}.

Cooperation of MT-based and actin-based transportation networks also 
play crucial roles in neuronal transport. Actin filaments help in 
the transport of the cargo to bridge the gaps between MT filaments 
\cite{brown}. The linkers of the two networks and the mechanisms of 
regulation and control of the proper switchover of cargo from one 
system to the other have received some attention in the last few 
years \cite{fuchs1,fuchs2,rodionov}.

\section{Collective transport by kinesins: molecular motor traffic jam}

Most of the multi-motor phenomena we have considered in the preceeding 
section are restricted to sufficiently low densities where direct 
interaction of the cargoes did not occur. As the cargoes are always 
much bigger than the motors (in-vitro as well as in-vivo), direct 
steric interactions of the cargoes become significant when several 
cargoes are carried by sufficiently dense population of motors along 
the same track. Such situations are reminiscent of vehicular traffic 
where mutual hindrance of the vehicles cause traffic jam at sufficiently 
high densities. In analogy with vehicular traffic, we shall refer to 
the collective movement of molecular motors along a filamentary track 
as ``molecular motor traffic''; we shall explore the possibility of 
molecular motor traffic jam and its possible functional implications.

\subsection{TASEP-like minimal models of molecular motor traffic}

Most of the minimal theoretical models of interacting molecular motors 
\cite{lipo,frey,frey2,santen,popkov} utilize the similarities 
between molecular motor traffic on MT and vehicular traffic on highways 
\cite{css} both of which can be modelled by appropriate extensions of 
driven diffusive lattice gases \cite{sz,schuetz}. In such models the 
motor is represented by a ``self-propelled'' particle and its dynamics 
is formulated as an appropriate extension of the totally asymmetric 
simple exclusion process (TASEP) which is one of the simplest models 
of interacting driven particles. Many aspects TASEP and its extentions 
as well as generalizations have been discussed in detail in the articles 
by Stinchcombe \cite{stinch}, Kolomeisky \cite{kolo} and Barma \cite{barma} 
in this proceedings. The novel feature of these models, which distinguish 
these from TASEP are the Langmuir-like kinetics of adsorption and 
desorption of the motors. 

\begin{figure}[tb]
\begin{center}
\includegraphics[width=0.4\textwidth]{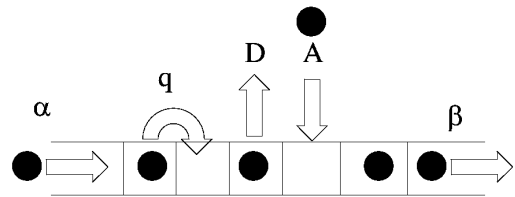}\\
\end{center}
\caption{Schematic description of the model of cytoskeletal motor traffic  
developed by Parmeggiani et al.\cite{frey}} 
\label{fig-frey}
\end{figure}

Let us consider the model suggested by Parmeggiani et al.\cite{frey}.  
In this model (see fig.\ref{fig-frey}) the self-propelled particle 
can hope from one lattice site to the next immediately in front of it, 
provided the target site is empty, with hopping probability $q$ per 
unit time. Moreover, a particle gets attached to an empty site 
with probability $A$ per unit time time, whereas a particle attached 
to a site can get detached with probability $D$ per unit time. 
Furthermore, the attachment probability at the entrance and the 
detachment probability at the exit are denoted by $\alpha$ and $\beta$, 
respectively, per unit time. In spite of the extreme simplicity of this 
model, it predicted a novel jammed phase on the $\alpha-\beta$ phase 
diagram. The progress made so far has been reviewed in this proceedings 
by Lipowsky et al.\cite{lipoetal}.

\subsection{\label{seckif}KIF1A traffic: effects of biochemical cycle}

\begin{figure}[tb]
\begin{center}
\includegraphics[width=0.4\textwidth]{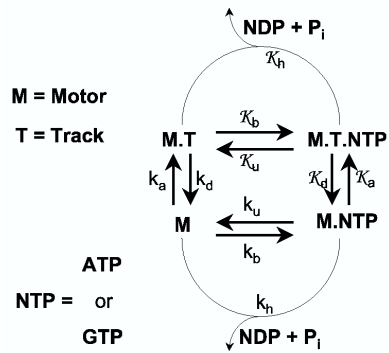}\\
\end{center}
\caption{Biochemical cycle in the absence and presence of the track. 
The symbols above (or below) the arrows are the rate constants 
for to the chemical reactions represented by the corresponding arrows.}  
\label{fig-bccycle}
\end{figure}

In reality, a molecular motor is an enzyme that hydrolyses ATP. 
In the presence of the track, usually the ATPase activity of the 
motor increases by one or two orders of magnitude; moreover,  
the mechanical movement of the motor is coupled to its biochemical 
cycle that includes its enzymatic activity (see fig.\ref{fig-bccycle}). 
Models developed for 
describing the mechanisms of operation of a single motor of a specific 
type incorporate not only the mechanical states of the motor but also 
its chemical states in each biochemical cycle. 
On the other hand, the models of interacting motors in traffic, 
which we mentioned in the preceeding subsection, do not 
explicitly take into account the distinct chemical states of each 
motor during its biochemical cycle.  Therefore, one common 
theme of our recent works on molecular motors  has 
been to achieve a synthesis- we incorporate the essential steps of the 
biochemical cycle, in addition to the mutual interactions of the motors, 
to develope models for their collective traffic-like dynamics. In the 
low-density limit, the model describes single-motor dynamics while the 
same model at higher densities predicts the collective spatio-temporal 
organization of the motors.

In a recent paper \cite{nosc} we considered specifically the
{\it single-headed} kinesin motor, KIF1A \cite{okada1,okada3,Nitta};
the movement of a single KIF1A motor had already been modelled with a
Brownian ratchet mechanism \cite{julicher,reimann}. In contrast to the
earlier models \cite{frey,frey2,santen,popkov,lipo} of molecular motor traffic,
which take into account only the mutual interactions of the motors, our
model (from now referred to as the NOSC model) explicitly incorporates 
also the Brownian ratchet mechanism of individual KIF1A motors, including 
its biochemical cycle that involves ATP hydrolysis.

\begin{figure}[tb]
\begin{center}
\includegraphics[width=0.4\textwidth]{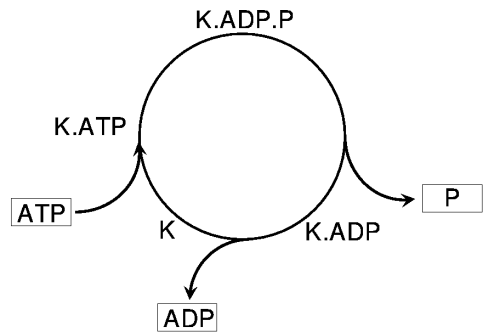}\\
\end{center}
\caption{The biochemical cycle of a single-headed kinesin motor KIF1A.}  
\label{fig-kif1acycle}
\end{figure}

The biochemical cycle of a single-headed kinesin motor KIF1A can 
be described by the four states shown in Fig.~\ref{fig-kif1acycle}
\cite{okada1,Nitta}: bare kinesin (K), kinesin bound with ATP (K.ATP),
kinesin bound with the products of hydrolysis, i.e., adenosine
diphosphate and phosphate (K.ADP.P), and, finally, kinesin bound with
ADP (K.ADP) after releasing phosphate. 

Through a series of in-vitro experiments, Okada, Hirokawa and
co-workers established that \\
(i) each KIF1A motor, while weakly bound to a MT, remains tethered
to the MT filament by the electrostatic attraction between the
positively charged $K$-loop of the motor and the negatively
charged $E$-hook of the MT filament. \\
(ii) In the weakly bound state, a KIF1A cannot wander far away from
the MT, but can execute (essentially one-dimensional) diffusive motion
parallel to the MT filament. However, in the strongly bound state,
the KIF1A motor remains strongly bound to a motor binding site on
the MT.\\
(iii) A transition from the strongly bound state to the weakly bound
state is caused by the hydrolysis of the bound ATP molecule. After
releasing all the products of hydrolysis (i.e., ADP and phosphate),
the motor again binds strongly with the nearest binding site on the MT.

In the NOSC model a single protofilament of MT is modelled by a 
one-dimensional lattice of
$L$ sites each of which corresponds to one KIF1A-binding site on the MT;
the lattice spacing is equivalent to $8$ nm which is the separation
between the successive binding sites on a MT \cite{howard}. Each kinesin
is represented by a particle with two possible internal states, labelled
by the indices $1$ and $2$, in which it binds strongly and weakly, 
respectively, to the MT track. Attachment of a motor to the MT occurs
stochastically whenever a binding site on the latter is empty.
Such a two-state approximation to the full sequence of biochemical 
states (conformations) has been successfully exploited also for 
conventional two-headed kinesin and unconventional myosin-V both 
of which are known to be processive \cite{kolo03,kolo05a}.

\begin{figure}[htb]
\begin{center}
\vspace{0.5cm}
\includegraphics[angle=-90,width=0.3\textwidth]{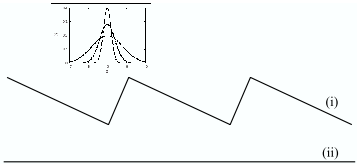}
\end{center}
\caption{Schematic description of the Brownian-ratchet. 
}
\label{fig-kifbrat}
\end{figure}

For the dynamical evolution of the system, one of the $L$ sites is
picked up randomly and updated according to the rules given below
together with the corresponding probabilities (Fig.~\ref{fig-kifbrat}):
\begin{eqnarray}
 &&{\rm Attachment:} \,\,\,\,\,  0\to 1 \,\,\,{\rm with} \,\, \omega_a dt\\
 &&{\rm Detachment:} \,\,\,\, 1\to 0 \,\,\, {\rm with} \,\, \ \omega_d dt\\
 &&{\rm Hydrolysis:} \,\,\,\,\,  1\to 2 \,\,\,{\rm with} \,\, \omega_h dt\\
 &&{\rm Ratchet:}\,\,\,\,\, \left\{\begin{array}{c}
     2 \to 1\,\,\,{\rm with} \,\, \omega_s dt\\
     20 \to 01\,\,\,{\rm with} \,\, \omega_f dt
   \end{array}\right.\\
 &&{\rm Brownian\ motion:}\,\,\,\,\, \left\{\begin{array}{c}
     20 \to 02\,\,\,{\rm with} \,\, \omega_b dt\\
     02 \to 20\,\,\,{\rm with} \,\, \omega_b dt
   \end{array}\right.
\end{eqnarray}

The physical processes captured by the rate constants $w_f$ and $w_s$
can be understood as follows by analyzing the Brownian ratchet mechanism
illustrated in fig.\ref{fig-kifbrat}. For the sake of simplicity, let
us imagine that the potential seen by the motor periodically oscillates
between the sawtooth shape and the flat shape shown in fig.\ref{fig-kifbrat}.
When the sawtooth form remains ``on'' for some time, the particle
settles at the bottom of a well. Then, when the potential is switched
``off'', the probability distribution of the position of the particle
is given by a delta function which, because of free diffusion in the
absence of any force, begins to spread. After some time the Gaussian
profile spreads to such an extent that it has some overlap also with
the well in front, in addition to the overlap it has with the original
well. At that stage, when the sawtooth potentil is again switched on,
there is a non-vanishing probability that the particle will find itself
in the well in front; this probability is proportional to the area of
that part of the Gaussian profile which overlaps with the potential 
well in front 
and is accounted for in our model by the parameter $w_f$. There is
also significant probability that the particle will fall back into
the original well; this is captured in our model by the parameter
$w_s$.
                                                                                
Let us denote the probabilities of finding a KIF1A molecule in
the states $1$ and $2$ at the lattice site $i$ at time $t$ by the
symbols $r_i$ and $h_i$, respectively. In mean-field approximation the
master equations for the dynamics of motors in the bulk of the system
are given by \cite{nosc}
\begin{eqnarray}
\frac{dr_i}{dt}&=&\omega_a (1-r_i-h_i) -\omega_h r_i -\omega_d r_i
+\omega_s h_i\nonumber\\
&&+\omega_f h_{i-1}(1-r_i-h_i),\\
\frac{dh_i}{dt}&=&-\omega_s h_i +\omega_h r_i
-\omega_f h_i (1-r_{i+1}-h_{i+1}) \nonumber\\
&&-\omega_b h_i (2-r_{i+1}-h_{i+1}-r_{i-1}-h_{i-1})  \nonumber\\
&&+\omega_b (h_{i-1}+h_{i+1})(1-r_i-h_i).
\label{eq-bulk}
\end{eqnarray}
The corresponding equations for the boundaries, which depend on the rate
constants of entry and exit at the two ends of the MT, are similar and 
will be presented elsewhere \cite{greulich}.

\begin{figure}[tb]
\begin{center}
\includegraphics[width=0.4\textwidth]{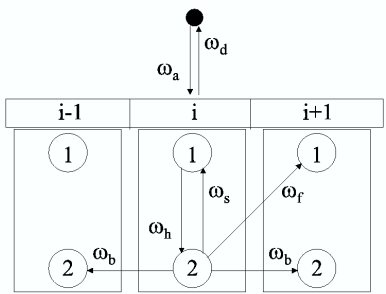}\\
\end{center}
\caption{A schematic representation of the NOSC model of KIF1A traffic. 
The integer $i$ denotes the discrete equispaced binding sites for 
kinesin motors on the MT track. The symbols $1$ and $2$ within circles 
at a given site $i$ represent the ``chemical'' (conformational) states 
of a kinesin at the same spatial location. }  
\label{fig-half}
\end{figure}

The single KIF1A properties are reproduced by this model \cite{nosc} in 
the appropriate low-density limit. For example, $v$, the mean speed of 
the kinesins, is about $0.2$~nm/ms if the supply of ATP is sufficient, 
and that $v$ decreases with the lowering of ATP concentration following 
a Michaelis-Menten type relation. 

Assuming {\em periodic} boundary conditions, the solutions
$(r_i, h_i)=(r,h)$ of the mean-field equations
(\ref{eq-bulk}) in the steady-state are found to be
\begin{eqnarray}
r&=&\frac{ -\Omega_h - \Omega_s -  (\Omega_s -1)K  +
{\sqrt{D}} }{2 K(1+K)},\label{eqr}\\
h&=&\frac{\Omega_h +\Omega_s + (\Omega_s +1)K
-{\sqrt{D}} }{2 K}\label{eqh}
\end{eqnarray}
where $K=\omega_d/\omega_a$,
$\Omega_h=\omega_h/\omega_f$, $\Omega_s=\omega_s/\omega_f$, and
\begin{equation}
 D=4\Omega_s K(1+K)+
{\left( \Omega_h +\Omega_s + ( \Omega_s-1)K  \right) }^2.
\end{equation}
The probability of finding an empty binding site on a MT is $Kr$ as
the stationary solution satisfies the equation $r+h+Kr=1$.
The corresponding flux is given by \cite{greulich}
\begin{eqnarray}
J &=& W\omega_f \nonumber \\
  &=& \frac{\omega_a\omega_h\omega_f}{\omega_f(\omega_a + \omega_d) + \omega_a(\omega_s+\omega_h) + \omega_d\omega_s} \nonumber \\
   &=& \frac{\omega_h(1+K)}{(1+K)[(1+K) + (\Omega_s + \Omega_h) + \Omega_s K]}.
\end{eqnarray}

We have also computed the average density profile of the motors along 
a MT track with open boundary conditions. For a given $\omega_a$, the 
density of motors in state 2 away from the edges of the MT exceeds 
that of those in state 1 as $\omega_h$ increases beyond a certain value.

\begin{figure}[htb]
\begin{center}
\includegraphics[width=0.45\columnwidth]{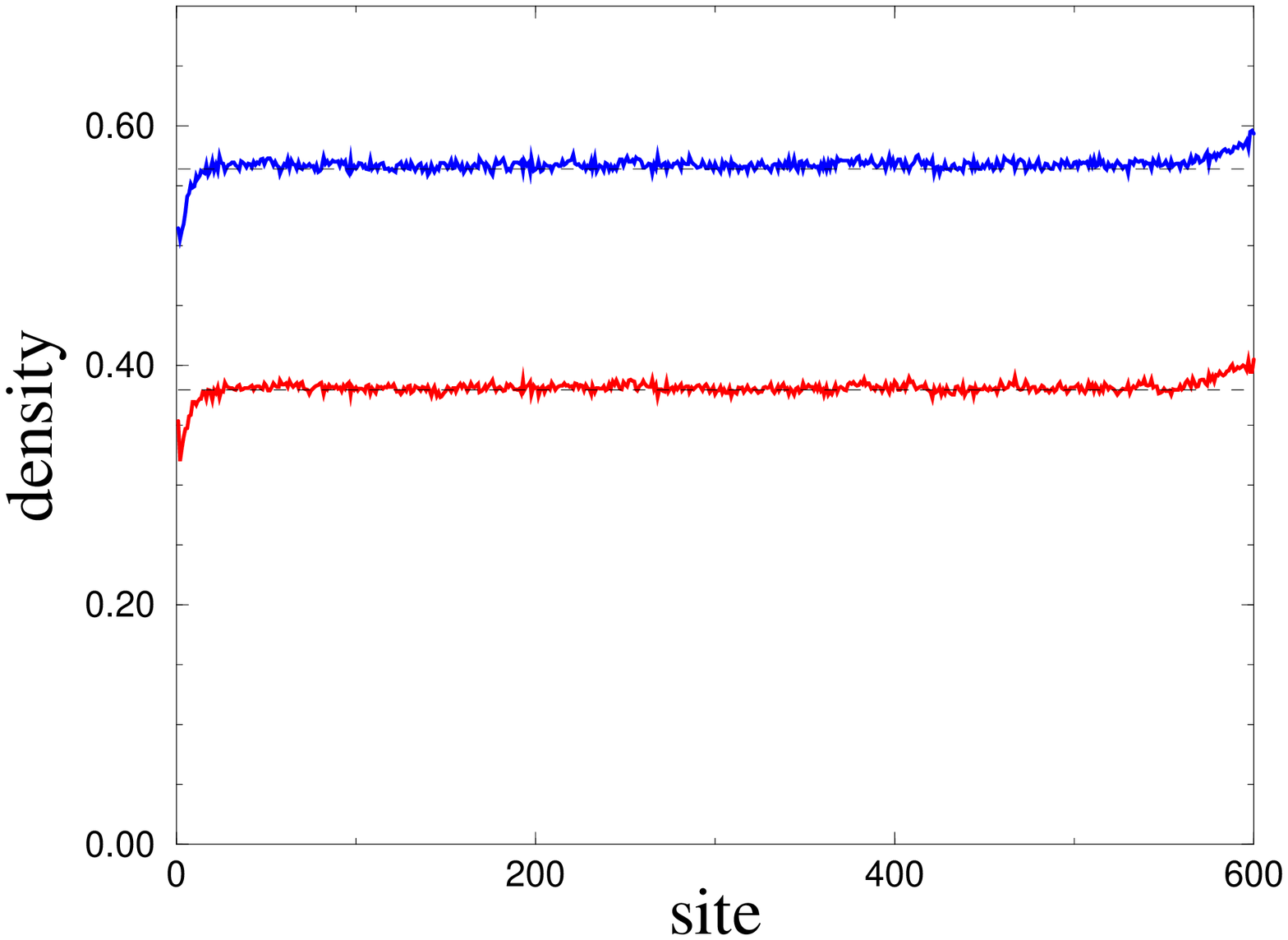}
\includegraphics[width=0.45\columnwidth]{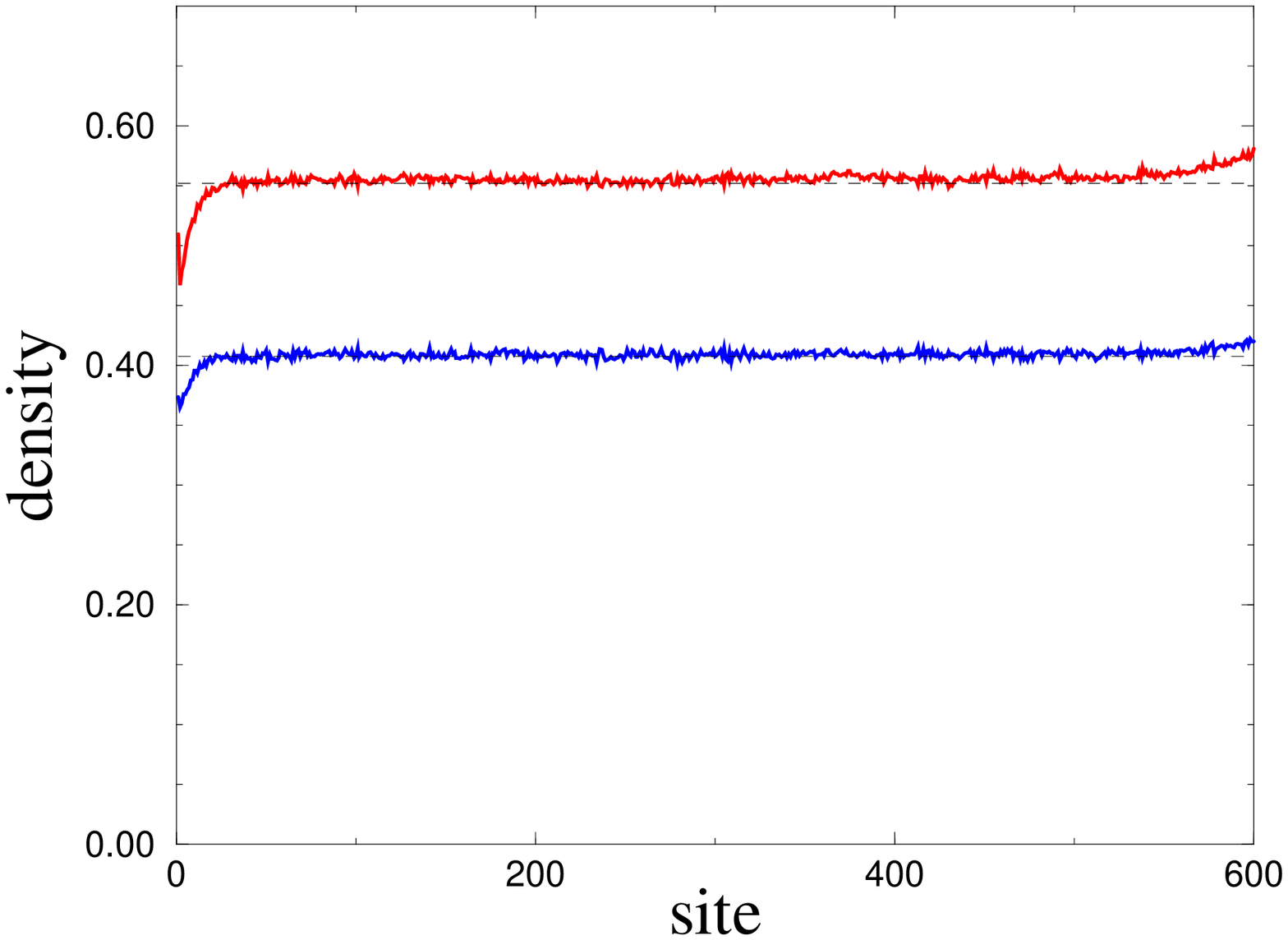}
\end{center}
\caption{The stationary density profiles for $\omega_h=0.1$ (left) and
$\omega_h=0.2$ (right) in the case $\omega_a=0.001$.
The blue and red lines correspond to the densities of state 1 and 2,
respectively. The dashed lines are the mean-field predictions 
for periodic systems with the same parameters.
}
\label{fig-density}
\end{figure}

In contrast to the phase diagrams in the $\alpha-\beta$-plane reported
by earlier investigators \cite{frey,santen,lipo}, we have drawn the
phase diagram of our model in the $\omega_a-\omega_h$ plane by carrying 
out extensive computer simulations for realistic parameter values of 
the model with open boundary conditions \cite{nosc,greulich}.
The phase diagram shows the strong influence of hydrolysis on
the spatial distribution of the motors along the MT. For very low
$\omega_h$ no kinesins can exist in state 2; the kinesins, all of which
are in state 1, are distributed rather homogeneously over the entire
system. However, immobile (but fluctuating) shock is observed in the 
densiti profiles of kinesins if $\omega_h$ is finite.

\section{\label{secribo} Ribosome traffic and protein synthesis }

{\it Translation}, the process of synthesis of proteins \cite{kapp} by 
decoding genetic
information stored in the mRNA, is carried out by {\it ribosomes}.
Understanding the physical principles underlying the mechanism of operation
of this complex macromolecular machine \cite{spirin02} will not only provide
insight into the regulation and control of protein synthesis, but may also
find therapeutic applications as ribosome is the target of many antibiotics
\cite{hermann}.
                                                                                
Most often many ribosomes move simultaneously on the same mRNA
strand while each synthesises a protein. In all the earlier models of
collective traffic-like movements of ribosomes
\cite{macdonald69,lakatos03,shaw03,shaw04a,shaw04b,chou03,chou04},
the entire ribosome is modelled as a single ``self-propelled particle''
ignoring its molecular composition and architecture. The typical size 
of an individual ribosome is much larger than that of a single codon 
(i.e., a triplet of nucleotides). 
This feature of the ribosome is captured in most of the recent 
theoretical models by postulating that the size of the self-propelled 
particle is $\ell$ ($\ell > 1$) where the unit of length is set by the 
size of a codon.  Moreover, in
these models the inter-ribosome interactions are captured through
hard-core mutual exclusion and the dynamics of the system is formulated
in terms of rules that are essentially straightforward extensions of
the TASEP \cite{schuetz}. The qualitative features of the flow 
properties and spatio-temporal organisation in these models are very 
similar to those of TASEP which corresponds to the special limit 
$\ell = 1$.

\begin{figure}[tb]
\begin{center}
\includegraphics[width=0.4\textwidth]{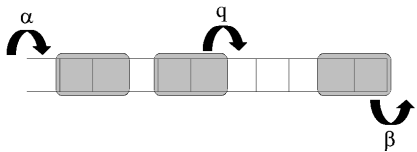}\\
\end{center}
\caption{Schematic description of the traffic of ribosomes where each 
ribosome has been treated as an ``extended'' particle ($\ell = 2$ 
in this figure.} 
\label{fig-ribotasep}
\end{figure}

\begin{figure}[tb]
\begin{center}
\includegraphics[width=0.4\textwidth]{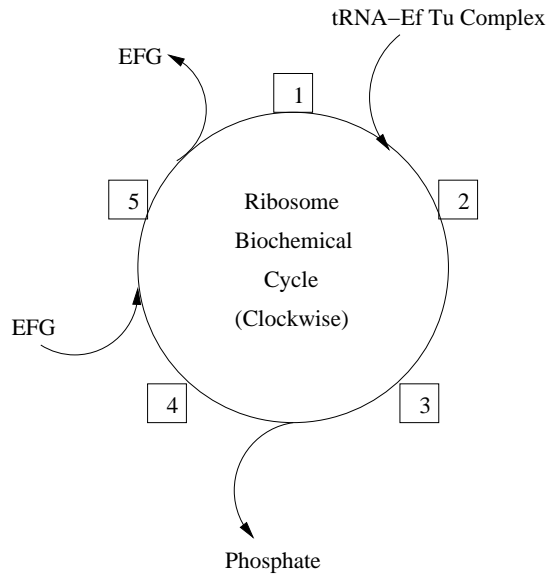}\\
\end{center}
\caption{A simplified biochemical cycle of s ribosome.} 
\label{fig-circle}
\end{figure}

In reality, a ribosome has a complex architecture. Each ribosome consists 
of two subunits and has four binding sites. Of these, three sites (called 
E, P, A), which are located in the larger subunit, bind to tRNA, while 
the fourth binding site, which is located on the smaller subunit, binds 
to the mRNA strand. The translocation of the smaller subunit of each 
ribosome on the mRNA track is coupled to the biochemical processes (see 
fig.\ref{fig-circle}) occuring in the larger subunit.

Let us begin the biochemical cycle with state 1 where a tRNA is bound to 
the site P of the ribosome. A tRNA-EF-Tu complex now binds to site A and 
the system makes transition from the state 1 to the state 2. As long as 
the EF-Tu is attached to the tRNA, codon-antiodon binding can take place, 
but the peptide bond formation is prevented. The EF-Tu has a GTP part 
which is hydrolized to GDP, driving the transition from state 2 to 3. 
Following this, a phosphate group leaves, resulting in the intermediate 
state 4. This hydrolysis, finally, releases the EF-Tu, and the peptide 
formation is now possible. After this step, EF-G, in the GTP bound form, 
comes in contact with the ribosome. This causes the tRNAs to shift from 
site P to E and from site A to P, site A being occupied by the EF-G, 
resulting in the state 5. Hydrolysis of the GTP to GDP then releases the 
EF-G followed by 
conformatinal changes that are responsible for pulling 
the mRNA-binding smaller subunit by one step forward. Finally, the tRNA 
on site A is released, resulting in completion of one biochemical cycle; 
in the process the ribosome moves forward by one codon \cite{alberts}. 

Very recently we \cite{basuchow} have developed a quantitative model 
that not only
incorporates the hard-core mutual exclusion of the interacting
ribosomes, but also captures explicitly the essential steps in the
biochemical cycle of each ribosome, including GTP (guanine
triphosphate) hydrolysis, and couples it to its mechanical movement
during protein synthesis. 
The modelling strategy adopted in ref.\cite{basuchow} for incorporating 
the biochemical
cycle of ribosomes is similar to that followed in our earlier work
\cite{nosc} on single-headed kinesin motors KIF1A. However, the
implementation of the strategy is more difficult in case of 
ribosome traffic not only
because of the higher complexity of composition, architecture and
mechano-chemical processes of the ribosomal machinery but also
because of the {\it heterogeneity} of the mRNA track \cite{nelson}.
The details of our work on ribosome traffic has been reported 
elsewhere \cite{basuchow}.

\section{Summary and conclusion}

In this paper we have presented examples of collective behaviours of 
molecular motors that emerge from the coordinations, cooperations 
and competitions at different levels in the sub-cellular world. 
In our original works so far we have synthesized the single-motor 
mechano-chemistry and multi-motor interactions to develope theoretical 
models that make experimentally testable quantitative predictions. 
In particular, we have developed models of KIF1A traffic on a MT 
track and ribosome traffic on a mRNA track.


\noindent{\bf Acknowledgements}: It is my great pleasure to acknowledge 
enjoyable collaborations with Aakash Basu, Ashoke Garai, Philip
Greulich, Katsuhiro Nishinari, Yasushi Okada and Andreas Schadschneider. 
I thank Aakash Basu for a critical reading of the manuscript.
I also thank Otger Campas, Erwin Frey, Frank J\"ulicher, Stefan Klumpp, 
Anatoly Kolomeisky, Reinhard Lipowsky, Gautam Menon and Alex Mogilner 
for useful discussions over the last few years on various aspects of 
molecular motors.



\end{document}